\documentclass[12pt]{article}
\usepackage{epsfig}
\usepackage{amssymb,amsmath}
\usepackage{eucal}
\newsavebox{\uuunit}
\sbox{\uuunit}
    {\setlength{\unitlength}{0.825em}
     \begin{picture}(0.6,0.7)
        \thinlines
        \put(0,0){\line(1,0){0.5}}
        \put(0.15,0){\line(0,1){0.7}}
        \put(0.35,0){\line(0,1){0.8}}
       \multiput(0.3,0.8)(-0.04,-0.02){12}{\rule{0.5pt}{0.5pt}}
     \end {picture}}

\newcommand{\eq}{\begin{equation}}
\newcommand{\feq}{\end{equation}}
\newcommand{\eqn}{\begin{eqnarray}}
\newcommand{\feqn}{\end{eqnarray}}
\newcommand{\arr}{\begin{eqnarray*}}
\newcommand{\farr}{\end{eqnarray*}}

\newcommand{\DD}{{\cal D}}

\newcommand{\RR}{{\cal R}}
\newcommand{\VV}{{\cal V}}

\newcommand{\ra}{\rightarrow}

\newtheorem{note}{Remark}[section]
\newtheorem{lemma}{Lemma}[section]
\newtheorem{corollary}{Corollary}[section]
\newtheorem{proposition}{Proposition}[section]

\csname @addtoreset\endcsname{equation}{section}

\def\rmd{{\rm d}}

\def\bfone{\relax{\rm 1\kern-.35em 1}}

\begin{document}
\begin{titlepage}
\begin{flushright}
IFUM-798-FT \\
KUL-TF-04/15 \\
\  \\
hep-th/0405284
\end{flushright}
\vspace{.3cm}
\begin{center}
\renewcommand{\thefootnote}{\fnsymbol{footnote}}
{\Large \bf New constraint for Black Holes  in ${\cal N}=2$, $D=5$ supergravity with matter} \vskip 15mm {\large
\bf {Sergio~L.~Cacciatori$^1$\footnote{cacciatori@mi.infn.it}
and Alessio~Celi$^2$\footnote{Alessio.Celi@fys.kuleuven.ac.be}}}\\
\renewcommand{\thefootnote}{\arabic{footnote}}
\setcounter{footnote}{0}
\vskip 10mm
{\small
$^1$ Dipartimento di Matematica dell'Universit\`a di Milano,\\
Via Saldini 50, I-20133 Milano, Italy\\

\vspace*{0.5cm}

$^2$  Instituut voor Theoretische Fysica, Katholieke Universiteit Leuven,\\
       Celestijnenlaan 200D B-3001 Leuven, Belgium.

\vspace*{0.5cm}

}
\end{center}
\vspace{2cm}
\begin{center}
{\bf Abstract}
\end{center}
{\small  With the general aim to classify BPS solutions in ${\cal N}=2, \quad D=5$ supergravity with
hypermultiplets and vector multiplets, here we consider a family of static spacetime metrics containing black
hole-like solutions, with generic hypermultiplets coupled to radially symmetric electrostatic vector multiplets.
We derive the general conditions which the fields must satisfy and determine the form of the fixed point
solutions. }
\end{titlepage}

\section{Introduction}

In this paper we generalize what it has been done in \cite{noialtri}, with the introduction of an arbitrary
number of vector multiplets considering only abelian gauge groups (${U(1)}^{n_V+1}$). This extension allows to
consider new classes of configurations due to the richer structure that characterizes the potential
\cite{ceresole:2001}. From a more general point of view this extension can be useful to understand which new
features arise when considering charged solutions in presence of both hypermultiplet and vector multiplet
couplings. A prominent role in this class is taken by the black hole configurations \cite{Behrndt}. They have
been in the recent years (\cite{noialtri} and references therein) and continue to be \cite{Ooguri:2004zv} an,
apparently never-ending, source of new insights in string theory. For this reason a lot of efforts have been
produced in the classification of BPS solutions \cite{classi} and in particular of black holes in the ungauged
and gauged supergravity coupled with vector multiplets in AdS$_5$ background \cite{Gutowski:2004bj},
\cite{Gutowski:2004ez}, \cite{Gutowski:2004yv}.

The work is organized as follows. We start presenting the model and the main ingredients of the theory
\cite{ceresole:2000}. We derive the integrability conditions for this case using the same ansatz  of
\cite{noialtri} for the metric and for the gauge fields and we study them together with the hyperini and gaugini
equations. As for the flat domain configurations of \cite{ceresole:2001} we obtain that the BPS conditions
ensure the stability of the potential as shown in \cite{stability}, \cite{stab2} and that the supersymmetric
flow equations are controlled by the superpotential $W$.

Although  the set of differential equations we get seems analogous to the sub-case $n_V=0$ treated in the our
previous work, a totally new feature arises. Indeed we find also for the scalars of very special geometry the
behavior $\varphi^{\prime \Lambda}\propto
\partial_\Lambda W$ (where with this notation we indicate generically all the scalars) as in
\cite{ceresole:2001},\footnote{As it will be emphasized later on, the factor of proportionality is not longer
the same for vector multiplet and hypermultiplet scalars.} \cite{daga}.

But in addition the special geometry imposes a sort of consistency constraint on the space-dependence of the
scalars of the vector multiplets. The consequences of such constraint are quite relevant: for example in the
$n_V=1$ case this determines completely the shape of the scalar for {\it any choice} of the gauging. The
analysis of this constraint (sec. \ref{ex_sol}) together with a preliminary study of the fixed point solutions
(sec. \ref{s3}) is a necessary conditions to construct a black hole configuration.

We explicitly show that the BPS conditions satisfy the equations of motion (to not tire the reader the
calculations are given in the appendix). At the end we conclude discussing the consequences and the possible
applications of our results.

\section{The model and its BPS equations}\label{model_v}
We consider $\cal{N}$=2 supergravity in five dimensions with an arbitrary number of hypermultiplets and vector multiplets.
The field content of the theory is the following:
\begin{itemize}
\item the supergravity multiplet
\eqn
\{ e^a_\mu \ , \psi^{\alpha i}_\mu \ , A^0_\mu \}
\feqn
containing the {\em graviton} $e^a_\mu$, two {\em gravitini}
$\psi^{\alpha i}_\mu$ and the {\em graviphoton} $A^0_\mu$;
\item $n_H$ hypermultiplets
\eqn
\{ \zeta^A \ , q^X \}
\feqn
containing the {\em hyperini} $\zeta^A $ with $A=1,2,\dots,2n_H$, and the
{\em scalars} $q^X$ with $ X=1,2,\dots,4n_H$ which define a quaternionic
Kahler manifold with  metric $g_{XY}$;
\item the vector multiplet
\eqn
\{ A^{\hat{I}}_\mu\ , \lambda^{ai} \ , \phi^x \}
\feqn
containing $n_V$ {\em gaugini} $\lambda^{ia} \ , \quad a=1, \ldots ,n_V$ with
spin $\frac 12$, $n_V$ real scalars $\phi^x $,
$ \quad x=1, \ldots ,n_V$
which define a very special manifold and $n_V$ {\em gauge vectors}
$A^{\hat{ I}}_\mu $ , $\quad \hat{I} =1\ldots , n_V $. Usually the graviphoton is included
by taking $I=0\ldots n_V$.
\end{itemize}

The bosonic sector of the gauged Lagrangian density is given by
\cite{ceresole:2000}
\eqn
{\cal {L}}_{BOS} &=& \frac 12 e\{ R-\frac 12 a_{IJ} F^I_{\mu \nu} F^{J \mu \nu}
-g_{XY} D_\mu q^X D^\mu q^Y -g_{x  y} D_\mu \phi^x D^\mu \phi^y
-2g^2 {\cal V} (q,\phi) \} \cr
& & +\frac 1{6\sqrt 6} \epsilon^{\mu\nu\rho\sigma\tau}
C_{IJK} F^I_{\mu\nu} F^J_{\rho\sigma}A^K_{\tau}
\label{lagr_v}
\feqn
with
\begin{gather*}
D_\mu q^X=\partial_{\mu} q^X + g A^I_\mu K_I^X(q) \\
D_\mu \phi^x=\partial_{\mu} \phi^x  + g A^I_\mu K_I^x(\phi)
\end{gather*}
where $K_I^X(q)$, $K_I^x(\phi)$ are the  Killing vectors on the quaternionic and the very special real manifold
respectively and ${\cal V} (q,\phi)$ is the scalar potential as given in Appendix.

At this point we concentrate our attention to the abelian case: this implies  that the action of the gauge group is
non trivial only on the quaternionic manifold while scalars of vector multiplet are uncharged under it.
 This means that $D_\mu \phi^x \equiv \partial_\mu \phi^x$ and   the existence of any isometry for the very special
 geometry is not required.
\noindent
Then the variations of the fermions for abelian gauge symmetry ${U(1)}^{n_V+1}$ reduce to:\\
 for the gravitini
\eqn \delta_\epsilon \psi_{\mu i} &=& \partial_\mu \epsilon_i +\frac 14 \omega^{ab}_\mu \gamma_{ab} \epsilon_i
-\partial_\mu q^X p_{Xi}^{\quad j} \epsilon_j + g A^I_\mu P_{Ii}^{\ j} \epsilon_j \cr &+&\frac i{4 \sqrt 6}
(\gamma_{\mu \nu \rho} -4g_{\mu \nu} \gamma_\rho )h_I F^{I\nu \rho } \epsilon_i - \frac i{\sqrt 6} g h^I
P_{Ii}^{\ j} \gamma_\mu \epsilon_j =0 \label{BPSgrav_v} \feqn

for gaugini \eqn \delta_\epsilon \lambda^x_i &=& \left[ -i \phi^{\prime x} e^{-w} \gamma_1 \delta_i^{\ j} -2 i g
h^{Ix} P_I^s (\sigma_s)_i^{\ j} \right. \cr & & \left. + \sqrt {\frac 32} e^{-w} h_I^x \left(v^\prime a^I +
{a^I}^\prime \right) \gamma_{01} \delta_i^{\ j} \right] \epsilon_j =0 \label{gaugini_v} \feqn

and for hyperini

\eqn \delta_\epsilon \zeta^A &=& f_{iX}^A \left[ - i q^{\prime X} e^{-w} \gamma_1 + i \sqrt {\frac 32} g a^I
K_I^X \gamma_0 +\sqrt {\frac 32}  g h^I K_I^X \right] \epsilon^i =0 \label{hyperini_v} \feqn where we have set
$\phi^{\prime x} :=\partial_r \phi^x$ and $q^{\prime X} :=\partial_r q^X$.\footnote{This notation applies to all
the quantities with the only exception of $a^I$ for which we explicitly define $a^\prime \equiv (\partial_r a^I)
h_I$. This choice is motivated by the aim to be manifest
 the similarities with the case $n_V=0$.}


As already explained at the beginning, we want to consider the direct generalization of the problem considered
in \cite{noialtri}. We look for electrostatic spherical solutions that preserve half of the ${\cal N}=2$
supersymmetries. We choose the same metric of the previous paper, which is $SO(4)$ symmetric with all the
other fields that only depend on the holographic space-time coordinate $r$. Moreover we fix the gauge for the
gauge fields keeping only the $A^I_t$ component different from zero.

Introducing  spherical coordinates $(t,r,\theta ,\phi, \psi)$ we write
 \eqn \label{metricansatz_v} ds^2 =-e^{2v} dt^2 +e^{2w}  dr^2 + r^2
(d\theta^2 +\sin^2 \theta d\phi^2 +\cos^2 \theta d\psi^2 ) \feqn
where the functions $v$ and $w$ depend on $r$ only.

We parametrize the vector fields as
\eqn \label{elettrico_v}
A^I_t =\sqrt {\frac 32}
a^I(r)e^v
\feqn
so that \eqn A_t^{\prime I} =\sqrt {\frac 32}
\left(v^\prime a^I+ {a^I}^\prime \right) e^v
\feqn

\subsection{Integrability conditions}\label{integ_v}
We now consider the BPS equations for the gravitini: their integrability
condition is the vanishing of their commutators; using the general
formulas in \cite{gen:to be published} one finds only four independent commutators
\eqn
& &\left\{ \left[ -\frac 12 \partial_r (v^\prime e^{v-w}) +\frac 12 e^{v-w}
(v^\prime a+ a^\prime )^2 +\frac {g^2}2 e^{v+w} W^2 \right] \gamma_0
\gamma_1 \delta_i^{\ j} \right. \cr
& & - \sqrt {\frac 32} ig   e^v (v^\prime a^I +{a^I}^\prime) P_I^s (\sigma_s )_i^{\ j}
- \sqrt {\frac 32} i g  e^v a^I \tilde{D}_r P_I^s (\sigma_s )_i^{\ j} -\frac g{\sqrt 6}
e^v \tilde{D}_r P^s (\sigma_s )_i^{\ j} \gamma_0 \cr
& & \left. -\left[ \frac i2 \partial_r (e^{-w} (v^\prime a+ a^\prime )) e^v
\delta_i^{\ j} +g^2 e^{v+w} a^I h^J P_I^r P_J^s \epsilon_{rs}^{\quad t}
(\sigma_t )_i^{\ j} \right] \gamma_1 \right\} \epsilon_j =0 \label{tr_v} \\
& & \cr
& & \cr
& & \left\{
\left[-\frac 12 v^\prime e^{v-2w} + \frac {g^2}2 r e^v W^2 \right] \gamma_0 \delta_i^{\ j}
   +\frac {ig}{\sqrt 6} r e^{v-w} (v^\prime a+ a^\prime) P^s (\sigma_s)_i^{\ j} \gamma_1 \right. \cr
& & + \left. \left[ \frac i2 e^{v-2w} (v^\prime a+ a^\prime) \delta_i^{\ j}
+g^2 r e^v a^I h^J P_I^r P_J^s \epsilon_{rst} (\sigma^t)_i^{\ j} \right]
\right\} \epsilon_j =0 \label{tteta_v}  \\
& & \cr
& & \cr
& & \left\{
\left[ \frac 12
\partial_r (e^{-w}) - \frac {g^2}2 r e^w W^2
\right] \gamma_1 \delta_i^{\ j}
+\frac i4 \partial_r \left[ r e^{-w} (v^\prime a+ a^\prime) \right] \gamma_0 \gamma_1  \delta_i^{\ j}
\right. \cr
& & \left.
-\frac {ig}{\sqrt 6} r (v^\prime a+ a^\prime) P^s (\sigma_s)_i^{\ j} \gamma_0
+\frac {ig}{\sqrt 6} r \tilde{D}_r P^s (\sigma_s)_i^{\ j}
\right\} \epsilon_j =0
\label{rteta_v} \\
& & \cr
& & \cr
& &
\left\{ \frac 12 \left[ 1- e^{-2w} -\frac {r^2 e^{-2w}}4
(v^\prime a+ a^\prime)^2 +\frac 23 g^2 r^2 W^2 \right]
\delta_i^{\ j}  +\frac i2 r e^{-2w} (v^\prime a+ a^\prime) \delta_i^{\ j} \gamma_0 \right. \cr
& & \left. +\frac {ig}{\sqrt 6} r^2 e^{-w} (v^\prime a+ a^\prime) P^s (\sigma_s)_i^{\ j} \gamma_0 \gamma_1 \right\}
\sin \theta \epsilon_j =0 \label{tetafi_v}
\feqn
with the scalar derivative defined as
$$
\tilde{D}_\mu = \partial_\mu \phi^{x} \partial_x +  \partial_\mu q^{ X} D_X
$$
Here we have defined $a :=h_I a^I$ and $a^\prime := h_I {a^I}^\prime$.

\subsection{Matter field conditions: Hyperini equation}\label{hypequa_v}
Now we compare the information coming from the integrability condition with the supersymmetric variation of the matter
fermions. We consider first the equation for the hyperini. Assuming that
\eqn
re^{-2w} (v^\prime a+ a^\prime) \neq 0
\label{cond_v}
\feqn
we can rewrite (\ref{tetafi_v}) in the form
\eqn \label{ansatz_v}
(if^0 \gamma_0 \delta_l^{\ k} +f^r (\sigma_r )_l^{\
k} \gamma_1 )\epsilon_k =\epsilon_l
\feqn
with
\eqn
& & f^r= - g r e^w W Q^r \label{fr_v} \\
& & f^0=-\frac {1- e^{-2w} - {\left(\frac {r e^{-w}}2 (v^\prime a+ a^\prime)\right)}^2
+g^2 r^2 W^2}{r e^{-2w} (v^\prime a+ a^\prime)} \label{fo_v}
\feqn
Also if we define $\Lambda =-f^0$ we have
\eqn
f^r =\pm \sqrt {1-\Lambda^2} Q^r  \label{parallela_v}
\feqn
Put in (\ref{hyperini_v}) and using (\ref{fquadrato}) we obtain
\eqn
[iA\delta_l^{\ k} +B^s (\sigma_s )_l^{\ k} ]\gamma_1 \epsilon_k =
[C \delta_l^{\ k} +iD^s (\sigma_s )_l^{\ k}] \epsilon_k
\feqn
with
\eqn
& & A=+\frac 12 q^{\prime Z} e^{-w} f^0 +\sqrt {\frac 32} ga^I D^Z P_I^s f_s \\
& & B^s =+\frac 12 \sqrt {\frac 32} ga^I K_I^Z f^s -q^{\prime X}
R^{sZ}_{\ \ \ X} e^{-w} f^0 -\sqrt {\frac 32} ga^I D^Z P_r f_t
\epsilon^{rts} \\
& & C=\frac 12 \sqrt {\frac 32} a^I K_I^Z g +\frac 12 \sqrt {\frac 32}f^0 gK^Z \\
& & D^s =\sqrt {\frac 32} ga^I D^Z P_I^s +g\sqrt {\frac 32} h^I D^Z P_I^s f^0 \feqn It is now easy to see that
this condition is not compatible with (\ref{ansatz_v}) so that one must put $A=B^s =C=D^s =0$ that
is\footnote{It is possible to obtain immediately the same result applying the general analysis of the hyperini
equation in \cite{tesi}.} \eqn
& & q^{\prime Z} e^{-w} \Lambda =\sqrt 6 ga^I D^Z P_I^s f_s \label{111_v}\\
& & q^{\prime X} R^{sZ}_{\ \ \ X} e^{-w} \Lambda = -\frac 12 \sqrt {\frac 32}
ga^I K_I^Z f^s +\sqrt {\frac 32} ga^I D^Z P_{I r} f_t \epsilon^{rts} \label{222_v}\\
& & a^I K_I^Z =K^Z \Lambda \label{333_v} \\
& & a^I D^Z P_I^s =h^I D^Z P_I^s \Lambda \label{444_v}
\feqn
Using (\ref{333_v}) and (\ref{444_v}) in (\ref{111_v}) and (\ref{222_v}) we find
\eqn
& & q^\prime_Z =\pm 3ge^w \sqrt {1-\Lambda^2} \partial_Z W \label{erreeffe_v}\\
& & q^{\prime X} R^{sZ}_{\ \ \ X} e^{-w} =
\mp \sqrt {1-\Lambda^2} \sqrt {\frac 32} g \left( \frac 12 K^Z Q^s
+\sqrt {\frac 32}W D^Z Q_r Q_t \epsilon^{rst} \right) \cr
&&\label{555_v}
\feqn
After contraction of (\ref{555_v}) with $K_Z$ we obtain
\eqn
\sqrt {\frac 32} g^2 r e^{2w} W Q^r
= -2\frac{q^{\prime X} D_X P^r}{{|K|}^2} \label{210_v}
\feqn
which gives
\eqn
q^{\prime X} D_X Q^r =0 \label{costante_v}
\feqn
\eqn
ge^w {|K|}^2 \sqrt {1-\Lambda^2} =\pm 2q^{\prime X} \partial_X W \label{corol_v}
\feqn

Also (\ref{fo_v}) and (\ref{fr_v}) can be rewritten as
\eqn
& & \sqrt {1-\Lambda^2} =\mp g r e^w W  \label{1fr_v} \\
& & \Lambda=
\frac {1- e^{-2w} - {\left(\frac {r e^{-w}}2 (v^\prime a+ a^\prime)\right)}^2
+g^2 r^2 W^2}{r e^{-2w} (v^\prime a+ a^\prime)}
\label{1fo_v}
\feqn
Using (\ref {1fr_v}) in (\ref{1fo_v}) we obtain
\eqn
1=\Lambda^2 e^{-2w}\left[ 1 +\frac {r}{2\Lambda} (v^\prime a+ a^\prime)\right]^2  \label{511_v}
\feqn
If we use (\ref{corol_v}) in (\ref{555_v}), the last one becomes
\eqn
K^2 q^{\prime X} R^s_{ZX} =- q^{\prime X} \partial_X W \left(\sqrt {\frac 32}  Q^s K_Z + 3W Q^t D_Z Q^r
\epsilon_{tr}^{\quad s}\right)  \label{4restrizione_v}
\feqn

Many other relations, which will be useful to check the equations of motion,
follow from (\ref{erreeffe_v}), (\ref{costante_v}), (\ref{corol_v}) and
(\ref{4restrizione_v}):
\eqn
& & {|K|}^2 |q^\prime |^2 =6\left(q^{\prime X} \partial_X W\right)^2 \label{1restrizione_v} \\
& & |q^\prime |^2 K_Z =2\sqrt 6 \delta_{rs} q^{\prime X} R^r_{XZ} Q^s
q^{\prime Y} \partial_Y W  \label{2restrizione_v} \\
& & K^Z =2\sqrt 6 \delta_{rs} Q^r R^{sXZ} \partial_X W \label{3restrizione_v} \\& & |q^\prime |^2 =
 \frac 32 {|K|}^2 g^2 e^{2w} (1-\Lambda^2 ) \label{5restrizione_v}\\
& & K_Z =\sqrt 6 W R^{rX}_{\quad Z} D_X Q_r \label{6restrizione_v} \\
& & |\partial W|^2 =\frac {{|K|}^2}6 \label{7restrizione_v} \\
& & 3W q^{\prime X} \partial_X W D_Z Q_t ={|K|}^2 Q^r q^{\prime X} R^s_{ZX} \epsilon_{srt}
\label{8restrizione_v}
\feqn

\subsection{Matter field conditions: Gaugini equation}\label{gauequa_v}
Next let us consider gaugini: using (\ref{ansatz_v}) to replace $\gamma_0\epsilon$ in (\ref{gaugini_v}) one easily
obtains
\eqn
& & \Lambda \phi^{\prime x}  + \sqrt {\frac 32} h_I^x
 (v^\prime a^I+ a^{\prime I}) =0 \label{1_v} \\
& & 2g \Lambda h^{x I} P_I^s - \sqrt {\frac 32} e^{-w} h_I^x
 (v^\prime a^I+ a^{I\prime}) f^s =0 \label{3_v}
\feqn
which gives
\eqn
\phi^{\prime x} f^s  =-2 g e^w  h^{x I} P_I^s
\feqn
and using
\eqn
h_x^{I} P_I^s =-\frac 32 \partial_x (W Q^s )
\feqn
one finally has
\eqn
& & \partial_x Q^s =0 \label{chiq_v} \\
& & \pm \sqrt {1-\Lambda^2} \phi^{\prime x} =3 g e^w  g^{x y}
\partial_y W \label{strana_v} \\
& & \sqrt 6 g\Lambda \partial_x W =\mp e^{-w} h_I^x (v^\prime a^I
+a^{\prime I}) \sqrt {1-\Lambda^2} \label{bene_v}
\feqn

We can rewrite the information on the scalars in a compact way defining
$$
\varphi^\Sigma =
\begin{cases}
\phi^x & \text{for } \Sigma=1,...,n_V \\
q^X       & \text{for } \Sigma=n_V+1,...,n_V+4n_H
\end{cases}
$$
as
\begin{gather}
\tilde{D}_r Q^s =0 \label{genq_v} \\
\varphi^{\prime \Lambda} =\pm 3 g e^w \left(1-\Lambda^2\right)^{\frac 12
\Delta} g^{\Lambda \Sigma} \partial_\Sigma W \label{genscalar_v}\\
\intertext{with}\\
\Delta=
\begin{cases}
- 1 & \text{for } \Lambda=1,...,n_V \\
  1 & \text{for } \Lambda=n_V+1,...,n_V+4n_H
\end{cases}
\end{gather}
where $g^{\Lambda \Sigma}$ is simply the product metric.

Let us discuss the consequences of the above relations. They are a generalization of the ones obtained in \cite{noialtri}.
First of all a strong similarity with the domain wall
case \cite{ceresole:2001} emerges again: this observation is non trivial because the two configurations are
quite different and it suggests that it should be possible to obtain a very general insight on BPS solutions in
presence of generic matter couplings. To be more specific in the two situations it happens that the phase of
prepotential $Q^r$ do not depend on the vector multiplet scalars: under this condition the potential ${\cal
V}(q,\phi)$ reduces the form that has been put forward for gravitational stability \eqn {\cal V}= -6W^2 + \frac
92 g^{\Lambda \Sigma} \partial_\Lambda W \partial_\Sigma W \feqn It is easy to see that in this case critical
points of W are also critical points of ${\cal V}$. Furthermore we find that $\varphi^\Lambda \propto
\partial_\Lambda W$ but now the gauge interaction distinguishes between charged $q^X$ and uncharged $\phi^x$ via
the factor $1-\Lambda^2$. At the end we want to underline the importance of (\ref{1_v}) that practically gives
the component of the field strength on $h_x^I$ and with (\ref{a_v}) determines it as a vector of special
geometry. This information will be crucial to check whether BPS solutions satisfy the equations of motion.

\subsection{Further restrictions}

As usual we have to compare the previous information with the one coming from the other integrability conditions.
Let us consider
equation (\ref{tteta_v}): it is easy to show that or all the coefficients
vanish or it must be equivalent to (\ref{ansatz_v}).
The first case reduces to the case in which all the
coefficients of (\ref{tetafi_v}) vanish.
The second case occurs when the following conditions are true:
\eqn
& & f^0 =-\frac {v^\prime  -g^2 r e^{2w} W^2}{v^\prime a+ a^\prime} \label{questa_v} \\
& & f^r=- gre^w W Q^r
\label{verificare_v}
\feqn
with $a^I P_I^r$  parallel to $h^J P_J^r$
\eqn
a^I P_I^r =\beta(r)h^J P_J^r  \label{parallelismo_v}
\feqn
for some function $\beta$.

 From the properties of very special geometry
 the modulus of vector $h_I$ can
be normalized to one $h_I h^I = 1$, so that the set $(h^I,h_x^I)$ is a base for the $n_V+1-$dimensional space
with $h_I h_x^I=0 $. Then the following relation holds
\begin{gather}
a^I = a h^I + l^x h_x^I \label{decomp_v}
\end{gather}

Using the above decomposition in (\ref{444_v}) and (\ref{parallelismo_v}) we get
\begin{gather}
\begin{cases}
(a-\Lambda)D_Z P^r = - l^x D_Z P_x^r \\
(a-\beta) P^r      = - l^x P_x^r
\end{cases}
\end{gather}
that taking in account the BPS demand $\partial_x Q^r = 0$ gives
\begin{gather}
\begin{cases}
\beta=\Lambda\\
a-\Lambda = \sqrt{\frac 32} l^x\partial_x \ln W = -\frac r{\sqrt{6}} l_x \phi^{\prime x}\\
a-\Lambda = \sqrt{\frac 32} l^x\partial_x \ln \partial_Z W
\end{cases}
\label{set_v}
\end{gather}

We continue to derive the other equations from integrability conditions.

Equation(\ref{questa_v}) together with (\ref{fo_v}) gives
\eqn
1+2g^2 r^2 W^2 +\frac {r^2 e^{-2w}}4 \left[v^{\prime 2} -(v^\prime a+ a^\prime)^2 \right]
= e^{-2w}(1+\frac {r}2 v^\prime )^2 \label{serve?_v}
\feqn
If we substitute (\ref{1fr_v}) into (\ref{questa_v})
\eqn
\Lambda= \frac {rv^\prime - 1 + \Lambda^2}{r(v^\prime a+ a^\prime)} \label{a_v}
\feqn
Using (\ref{ansatz_v}) in (\ref{rteta_v}) we obtain the equations
\eqn
& & g r (v^\prime a+ a^\prime) W +g r \Lambda W^\prime \mp \frac 12 \partial_r
\left[r e^{-w}(v^\prime a+ a^\prime)\right]\sqrt {1-\Lambda^2}=0
\label{b_v} \\
& & \mp g r (v^\prime a+ a^\prime) W \sqrt {1-\Lambda^2} +\Lambda\partial_r
(e^{-w}) - \Lambda g^2 r e^w W^2 \cr
& & +\frac 12 \partial_r \left[r e^{-w} (v^\prime a+ a^\prime)\right] =0
\label{c_v}
\feqn
Similarly from (\ref{tr_v}) we have
\eqn
& & \Lambda g\tilde{D}_r \left( \sqrt {\frac 32} e^v a^I P_I^s \right) +\frac g2 e^v
W^\prime Q^s
\mp \sqrt {1-\Lambda^2} Q^s \left[ \frac 12 g^2 e^{v+w} W^2 + \frac 12 e^{v-w}
\left(v^\prime a + a^\prime \right)^2 \right. \cr
& & \left. -\frac 12 \partial_r (v^\prime e^{v-w}) \right] =0 \label{d_v}
\feqn

\eqn
& & \mp  \sqrt {1-\Lambda^2} ge^v W^\prime +\Lambda e^v \partial_r \left[e^{-w}
\left(v^\prime a + a^\prime \right) \right] +g^2 e^{v+w} W^2 +e^{v-w} \left(v^\prime a + a^\prime \right)^2
\cr
& & -\partial_r (v^\prime e^{v-w})=0 \label{e_v}
\feqn

Note that all the relations already derived reduce to those in \cite{noialtri} if we take $\Lambda = a$. So it
is easy to conclude that the particular case $l^x \equiv 0$ is compatible with the BPS conditions and reduces to
the set of equations (5.1)-(5.4) of \cite{noialtri} plus the one for the scalars of the vector multiplets.

\section{Static BPS configurations}\label{s3}

In this section we derive the independent set of equations that characterizes
BPS configurations.

We start by considering integrability conditions (\ref{b_v}) and (\ref{c_v}): subtracting from (\ref{b_v}) the
equation (\ref{c_v}) multiplied by $\mp \sqrt{1-\Lambda^2}$ we get \eqn gr \Lambda (v^\prime a + a^\prime) W +
gr W^\prime \mp \sqrt{1-\Lambda^2} w^\prime e^{-w} \mp g^2 r \sqrt{1-\Lambda^2} e^w W^2 = 0 \feqn Using
(\ref{a_v}) and (\ref{1fr_v}) it gives \eqn W^\prime = - (v^\prime + w^\prime) W \label{b-c_v} \feqn that
implies $gr_0 W= \mp e^{-(v+w)}$ where $r_0$ is a constant. This last expression can be rewritten considering
again (\ref{1fr_v}) as \eqn e^v= \frac r{r_0 \sqrt{1-\Lambda^2}} \label{delta_v} \feqn which is fundamental to
demonstrate the compatibility of the BPS conditions. Indeed taking the derivative with respect to $r$ and
comparing with (\ref{a_v}) we obtain \eqn v^\prime a + a^\prime = v^\prime \Lambda + \Lambda^\prime
\label{key_v} \feqn This means that the integrability conditions and consequently the BPS equations for the
metric ($w$ and $v$) and for the scalars of the hypermultiplets have the same form as the ones in
\cite{noialtri}: hence their consistency is ensured and the only change is the replacement of $a$ by $\Lambda$ .
The new ingredients here, due to the introduction of vector multiplets, are then the equation for $\phi'^x$ and
the relations between $a$, $l^x$ and $\Lambda$ (\ref{set_v}).

Note that here $a'$ is not the derivative of $a$ with respect to $r$. Indeed we have defined  $a' \equiv h_I
a^{\prime I}$. Using (\ref{decomp_v}) it is easy to find \eqn a' =\partial_r a -\sqrt {\frac 23} l_x
\phi^{\prime x} \feqn Substituting it in (\ref{key_v}) and using the second eq. of (\ref{set_v}) we get

\eqn v^\prime a + \partial_r a +\frac 2r (a-\Lambda) = v^\prime \Lambda + \Lambda^\prime \feqn

which gives

\eqn \Lambda=a + \frac {\mu}{r^2} e^{-v} \label{777} \feqn

where $\mu$ is an integration constant. According to (\ref{set_v}) the last expression can be rewritten  to show
in a transparent manner the relation between $\mu$ and $l^x$ as

\eqn \frac {\sqrt 6 \mu}{r^3} e^{-v} = l_x \phi^{\prime x} \feqn

The implication of this expression on the existence of fixed points still has to be clarified. Now it is not so
difficult to show that the BPS conditions we have derived satisfy the equations of motion. We refer the reader
to the appendix \ref{app:motion} for the technical details.

To summarize what it has been obtained, we conclude this section presenting a set of independent BPS equations

\begin{gather}
1={\Lambda}^2 e^{-2w} \left[ 1 +\frac {r}2
\left( v^\prime + \frac {{\Lambda}^\prime} {\Lambda} \right) \right]^2 \label{5.1_v} \\
e^v =\frac {r}{r_0 \sqrt {1-{\Lambda}^2}} \label{def_v_v}\\
\sqrt {1-{\Lambda}^2} =\mp g e^w rW \label{def_w_v}\\
{q^\prime}^Z =\pm 3ge^w \sqrt {1-{\Lambda}^2} \partial^Z W \label{5.4_v}\\
\phi^{\prime x} =\pm 3ge^w \frac 1{\sqrt {1-{\Lambda}^2}} \partial^x W \label{5.5_v}\\
\Lambda=a + \frac {\mu}{r^2} e^{-v} \label{5.6_v}\\
\frac {\sqrt 6 \mu}{r^3} e^{-v} = l_x \phi^{\prime x}\label{5.7_v}\\
\Lambda \phi^{\prime x}  + \sqrt {\frac 32} h_I^x
 (v^\prime a^I+ a^{\prime I}) =0 \label{5.8_v}
\end{gather}

We note that the last four equations are the new ones with respect to \cite{noialtri} due to the presence of
vector multiplets.

As an application we can immediately determine the fixed point solutions of these equations. Strictly speaking
they are the  solutions having constant scalars, that are defined by the conditions $\phi^{\prime x}=0$ and
$q^{\prime X} =0$. However one can include also the asymptotic fixed point solutions, which are characterized by
$\phi^{\prime x} \rightarrow 0$ and $q^{\prime X} \rightarrow 0$ for some special values of $r$.

For now we just consider the first case. First of all one finds that the fixed point solutions correspond to the
stationary point solutions  of the potential $W$: \eqn K_Z =\partial_Z W =\partial_z W=0 \ . \feqn
giving a fixed value $W\neq 0$. \\
Furthermore (\ref{5.7_v}) requires  $\mu =0$ so that $\Lambda=a$ whereas $Q^r$ must be
covariantly constant and finally the relation $h^{Ix} P^s_I =0$ must be true.\\
The resulting configuration is then
\eqn
& & e^v =\frac {\gamma +r^2}{r^2} \delta \sqrt {1+\frac {g^2 W^2 r^6}{(\gamma +r^2)^2}} \\
& & e^w =\frac {r^2}{\gamma +r^2} \frac 1{\sqrt {1+\frac {g^2 W^2 r^6}{(\gamma +r^2)^2}}}\\
& & F^I_{rt} =-\sqrt 6 \frac {\gamma \delta}{r^3} h^I \feqn where the two integration constants $\gamma$ and
$\delta$ are  related to the electric charges $Q^I$ by
 \eqn Q^I =\sqrt 6 \gamma \delta h^I \ . \feqn

 We have derived these solutions (as we have done for the BPS conditions (\ref{5.1_v}-\ref{5.8_v})) assuming
 $0<\Lambda^2<1$, it remains to study the two singular case
$\Lambda^2 =1$ and $\Lambda =0$.\footnote{It is immediate to see that no solutions with non trivial scalars
exist for $|\Lambda|=1,0$, so the present analysis covers all the possible configurations of this kind.} It is
quite easy to show that the last case doesn't correspond to any solution. For $\Lambda^2 =1$ the BPS equations
do not imply the equation of motion because some of them  become singular (and consequently the demonstration
given in the appendix \ref{app:motion} does not hold). However it is sufficient to impose by hand the Maxwell
equations. The resulting configuration is

\eqn
& & W=P_I=0  \\
& & e^v =c \left( 1-\frac b{2r^2} \right) \\
& & e^w =\frac 1{1-\frac b{2r^2}} \\
& & F^I_{rt} =\pm \sqrt {\frac 32} \frac {bc}{r^3} h^I
\feqn
This configuration corresponds to  the Reissner-Nordstr\"om (extreme) black-hole
of the minimal gauged theory and can be obtained from the general case as a
limit $W\rightarrow 0$. Actually from the comparison of the two cases it
seems that the presence of a nonvanishing $W$ ``regularizes'' the horizon
which disappears.

\par
At this point one should analyze the class of asymptotically fixed point solutions which can be obtained
perturbing the configurations just found. This requires a much more subtle investigation.

\section{Towards an explicit solution}\label{ex_sol}

As we have already argued in the introduction the most interesting solutions of the form (\ref{metricansatz_v})
are the ones which
correspond to asymptotically AdS extreme black holes. However to find solutions of this kind solving the BPS
equations (\ref{5.1_v}-\ref{5.8_v}) for an explicit model is quite hard. To understand better the origins of
these difficulties, let us consider the other (few) classes of BPS solutions with hypermultiplet and vector
multiplet couplings turned on already existing in the literature.
Also for very simple models it seems always necessary to use a numerical approach. So it can be easily supposed
that the numerical treatment will be the only possibility to perform explicit solutions. For example this
happens in the flat domains wall case \cite{ceresole:2001}. Although there are some similarities that we have
already discussed, our case is much more complicated by the presence of $\Lambda$. Indeed our equations become
almost of that form only for $\Lambda'=0$. But this choice is too restrictive because it fixes completely the
metric and, unlikely, to a form that does not have the nice features we are looking for \cite{noialtri}.

In addition, we have to satisfy (\ref{5.5_v}-\ref{5.8_v}) which, as we will discuss later on, can be seen as a
sort of consistency constraint. These ones together with presence of $\Lambda$  are a nontrivial obstruction to
the application of a numerical method without assuming any ansatz on the form of $\Lambda$ and to distinguish
the true solutions from the artifacts.

So let us discuss the general features of eq. (\ref{5.5_v}-\ref{5.8_v}) focusing in particular on the meaning of
the last three equations. Indeed these give the field strength $F^I$ in terms of $\phi'^x$ and $\Lambda$:

\eqn && F_{rt}^I \propto v' a^I + (a^I)' = (v' \Lambda + \Lambda') h^I -\sqrt{\frac 23} \Lambda \phi'^x h_x^I
\label{FS} \feqn

Now we have to compare this expression with the decomposition for $a^I$ (\ref{decomp_v}) and with (\ref{set_v}).
Following the analysis in the section \ref{s3} we obtain the relations regarding the $h^I$ projection that are
exactly the (\ref{5.6_v}) and (\ref{5.7_v}). At this point we have to study the consequences of the $h^I_x$
projection:

\eqn && \partial_r l^x + l^z \phi'^y B_{zy}^x -\sqrt{\frac 23} (a-\Lambda) \phi'^x + v' l^x= 0
\label{extra_v}\feqn

where we define $B_{zy}^x=B_{yz}^x \equiv (\partial_y h_z^J) h_J^x$. Together with

\eqn && a-\Lambda = -\frac r{\sqrt 6} l_x \phi'^x = -\frac {\mu }{r^2} e^{-v}  \label{extra_v1}\feqn

(\ref{extra_v}) imposes a non trivial constraint that the solution must satisfy. To clarify better the meaning
of this statement let us consider the specific case of one vector multiplet with a generic number of
hypermultiplets present (in fact this analysis can carry out for any other specific model in which $B_{zy}^x$ is
known explicitly). Using the parametrization in \cite{ceresole:2001} we have \eqn && B_{\rho\rho}^\rho = -\frac
32 \frac{(\partial_\rho^2 h^J) \partial_\rho h_J}{g_{\rho\rho}} = \frac 1\rho\feqn
 that gives

 \eqn
 && l= l_0 e^{v(r_0)} \frac {\rho(r_0)}{\rho(r)} \ \exp\left[-4\int_{r_0}^r dr r (\frac {\rho'}\rho)^2\right] e^{-v}
 \label{elle_v}\\
 &&  \frac {\rho'}{\rho^3} \ \exp\left[-4\int_{r_0}^r dr r (\frac {\rho'}\rho)^2\right] = \frac C{r^3}
 \label{extra_v2} \feqn
where $C= \frac{\sqrt 6 \mu e^{-v(r_0)}}{12 l_0 \rho(r_0)}$.
 For example, if we suppose for $\rho$ a power-law behavior, $\rho= \alpha r^\beta$, the above relation fixes
 $\beta$ to be or $-1$ or $1/2$ that rules out a lot of possible solutions. Indeed it is possible to show that these are
 the only non trivial solutions of (\ref{extra_v2}). To see this let us consider the derivative of (\ref{extra_v2}): this
 condition reduces to an ordinary differential equation

 \eqn
 && y' = 4 r y^3 + 2 y^2 -\frac 3r y
 \feqn

 where $y$ is the logarithmic derivative of $\rho$, $y\equiv \frac{\rho'}{\rho}$.
The above equation can be expressed in a convenient form (to be easily integrated by separation of variables) in
terms of $t\equiv y/z$ where $z=1/r$:

\eqn && \frac{dt}{dz} = -\frac 4z t(t+1)(t-\frac 12) \label{diffcond}\feqn

One recognizes immediately in the three constant solutions of (\ref{diffcond}), $t=0$, $t=-1$, $t=1/2$,
respectively the trivial solution $\rho'=0$ of (\ref{extra_v2}) and $\rho=\alpha r^\beta$, $\beta=-1,1/2$.

The non constant solutions live in the four regions delimited by the constant ones and they are defined by

\eqn && \left|\frac t{t_0} \right|^{1/2} \left|\frac{t+1}{t_0+1}\right|^{-1/6}
\left|\frac{t-1/2}{t_0-1/2}\right|^{-1/3} = \frac z{z_0}\feqn

At the end all the problem reduces to compute the real roots of a third-order polynomial

\eqn && (1-b) t^3 + \frac 34 b\ t - \frac b4= 0 \feqn

where $b=( c z)^6$, hence greater than zero, for $t>0 \vee t<-1$ while $b= - (c z)^6$ elsewhere. $c$ is a
positive constant of integration.  Performing explicit calculations, one obtains that all the solutions for
$t>0$ ($t<0$) behave asymptotically for $z\rightarrow \infty$ like $t=1/2$ ($t=-1$). The solutions  in the
regions $t<-1$ and $t>1/2$ exist only for a finite range in $z$,\footnote{Thus we can discard this kind of
solutions immediately.} $z>1/c$, while the solutions in the intermediate regions $0<t<1/2$ and $-1<t<0$
interpolate between $t=0$ and respectively $t=1/2$ and $t=-1$. At this point one has to check which of the
solutions of (\ref{diffcond}) satisfy also (\ref{extra_v2}). It can be easily shown, for example considering the
behavior for $z\simeq 0$, that only the solutions with $t$ constant survive.

Let us stress the relevance of the condition derived in this section. First of  all the appearance of such
strong requirement on the shape of the scalars of vector multiplets is quite surprising and, up to our
knowledge, new. In particular it seems quite striking that for $n_V=1$ the form of $\rho$ is fixed a priori for
any number of hypermultiplets and for any choice of the gauging. Indeed the implications of the above result on
the construction of electro-static spherically symmetric solutions are quite severe. As first consequence one
aspects that a solution of this kind {\it can not exist} for a generic selection of the isometries. Indeed this
is exactly the case: it is easy to check that for $n_H=1$ for any combination of the killing vector of the
form\footnote{We follow the notation and the parametrization of \cite{ceresole:2001} for the metric and the
isometries of the universal hypermultiplet, see the appendix \ref{convention}.}

\eqn &&K=h^0K_0 +h^1 K_1\\
&& K_0= \alpha k_{(1)} + \beta k_{(2)} \ \ \ \ \ \ \ \ \ K_1= \gamma k_{(1)} + k_{(2)} \label{tent}\feqn

with $\alpha$, $\beta$, $\gamma$ constant parameters. Due to the structure of the differential equations for the
scalars the same (non) result holds also substituting $k_{(2)}$ with $k_{(3)}$ in (\ref{tent}).

As second consequence, closely related to the first, it does not exist a general criteria to determine which are
the right choices for the gauging (in the sense that produce a solution) without an explicit try. Indeed, the
general procedure is to evaluate the equation (\ref{5.5_v}) for $\rho$, for the chosen prepotential  and
imposing the constraint. In this way one obtains an {\it algebraic } equation that the hypermultiplet scalars
have to satisfy. This way of acting is quite laborious and imposes a strong limitation on the number of models
that it can test.

 It could be nice to understand what happens in the presence of a generic number of vector multiplets. One
 expects that with more scalars  the requirement will be in some sense relaxed. Anyway the system should be also
 in this case overconstrained.
\section{Discussion}

In this section we want to recall the results already obtained and to point out which topics deserve more study.
First of all we have derived BPS equations, studying in the line of \cite{noialtri}, the relations from the
hyperini
 and the gaugini and the integrability conditions for the gravitini. We observe that the former ones have the same structure
 manifested in the domain wall case \cite{ceresole:2001}:  this suggests the possibility of determining some properties of
 BPS solutions without starting from the specific ansatz. The importance of a similar study is evident: for example this
 could permit us to give a definitive answer in the quest for a realistic cosmological model in gauged supergravity.

At the same time we have discovered a quite unexpected condition for the scalars of vector multiplets. As it
emerges quite clearly from the last section, the analysis and the better understanding of this constraint is
crucial for the construction of non trivial solutions. Let us stress again that for $n_V=1$ the relations
(\ref{elle_v}), (\ref{extra_v2}) are sufficient to determine the space-time dependence of the vector multiplet
scalar $\rho$ independently by the choice of the prepotential. This last observation suggests that it could be
possible to give an interpretations to this phenomena in terms of the six dimensional gauged supergravity where
the scalars of the vector multiplets are just a component of the gauge fields. Another interesting question that
arises quite naturally is whether  this kind of relations is peculiar to this particular case or instead is a
feature common to a larger class of charged solutions. This points are currently under investigation.

{\bf Acknowledgments}

We would like to thank F. Belgiorno, D. Klemm,  G. Smet, J. Van den Berg, A. Van Proeyen and D. Zanon for useful
discussions. This work is supported by the European Commission RTN program HPRN-CT-2000-00131 and partially
supported by INFN, MURST and the Federal Office for Scientific, Technical and Cultural Affairs through the
"Interuniversity Attraction Poles Programme -- Belgian Science Policy" P5/27.\normalsize

\newpage
\begin{appendix}

\section{Conventions}\label{convention}

In this appendix we present some definitions and properties that we use in our work. With \eqn
\label{quatscalars} q^X \, \qquad {\scriptstyle X}=1, \ldots , 4 n_H \feqn we denote the scalars of the
hypermultiplets which are the coordinates of a quaternionic manifold. We introduce the $4n_H$beins as \eqn
\label{fielbein} f^{iA}_X (q^Y ) \ ,\qquad i=1,2 \in SU(2) \ , A=1,\ldots,2n_H \in Sp(2n_H) \feqn

The splitting of the flat indices in $i$ and $A$ reflects the factorization of the holonomy group in
$USp(2)(\simeq SU(2))\otimes USp(2n_H)$ which is the main feature of those spaces. The indices as a consequence
of  the symplectic structure are highered and lowered with the antisymmetric matrices \eqn
&& \epsilon_{ij} \ , \qquad \mathbb{C}_{AB} \\
&& \epsilon_{ij}=\epsilon^{ij} \ , \qquad \epsilon_{12} =1 \\
&& \mathbb{C}_{AB} \mathbb{C}^{CB} =\delta_A^{\quad C} \ , \qquad \mathbb{C}^{AB} =(\mathbb{C}_{AB})^* \ . \feqn
following the NW-SE convention \cite{ceresole:2001}.

The important relation \eqn f_{XiC} f_{Yj}^{\ \ C} =\frac 12 \epsilon_{ij} g_{XY} + R_{XYij} \label{fquadrato}
\feqn can be viewed as  a definition for the quaternionic metric $g_{XY}$ and for the $SU(2)$ curvature
$R_{XYij}$.

We use the symbols $p_{Xi}^{\ \ \ j}$ for the $SU(2)$ spin connection whereas $\omega_\mu^{ab}$ denotes the
usual Lorentz spin connection. The covariant derivative which appears in the gravitini supersymmetry variation
acts on the  symplectic Maiorana spinors $\epsilon_i$ as \eqn \DD_\mu \epsilon_i  &=&
\partial_\mu \epsilon_i +\frac 14 \omega^{ab}_\mu \gamma_{ab} \epsilon_i -\partial_\mu q^X p_{Xi}^{\ \ j}
\epsilon_j - g A^I_\mu {P_I}_i^{\ j} \epsilon_j \feqn where  the generalized spin connection receives the
following contributions: the first term represents the Lorentz action while the others can be identified  with
the $SU(2)$ action plus a term due to the $SU(2)$ R-symmetry  gauging. $A^I_{\mu}$ are $(n_V+1)$ $1-$forms and
$P_I^r$ are the prepotentials while $g$ is the gauge coupling. We adopt the convention to define for the
quantities with an $I$ index the corresponding ``dressed'' ones like $P^r\equiv P_I^r h^I$ or $F^{\mu\nu}\equiv
F^{\mu\nu}_I h^I$. We note that in this notation the subcase $n_V=0$ is recovered in a natural way being $I=0$
and $h^I=h^0=1$.\footnote{This implies, from the definition of very special geometry, that the normalization of
$C_{I J K}$ is given by $C_{000}=1$.}

It is useful to introduce the projection on the Pauli matrices for quantities in the adjoint representation of
$SU(2)$, for example \eqn R_{XYi}^{\qquad j}=R_{XY}^r (i\sigma_r)_i^{\ j} \label{pro} \feqn where
$(\sigma_r)_i^{\ j}$ are the usual Pauli matrices \eqn \sigma_1 = \left( \begin{array}{cc} 0 & 1 \\ 1 & 0
\end{array} \right) \ , \qquad \sigma_2 = \left( \begin{array}{cc} 0 & -i \\ i & 0 \end{array} \right) \ ,
\qquad \sigma_3 = \left( \begin{array}{cc} 1 & 0 \\ 0 & -1 \end{array} \right) \ . \feqn which satisfy \eqn & &
(\sigma_r)_i^{\ j} (\sigma_s)_j^{\ k} =\delta_{rs} \delta_i^{\ k}
+i\epsilon_{rs}^{\quad t} (\sigma_t)_i^{\ k} \label{prodotto} \\
& & \left[\sigma_r ,\sigma_s \right] =2i \epsilon_{rst} \sigma^t \label{parentesi} \feqn The prepotentials are
defined by the relation \eqn
& & R_{XY}^r K^Y = D_X P^r \label{prepot} \\
& & D_X P^r := \partial_X P^r +2\epsilon^{rst} p_X^s P^t \feqn where $D_X$ is the $SU(2)$ covariant derivative.
They can be expressed in terms of the Killing vectors \eqn P^r =\frac 1{2n_H} D_X K_Y R^{rXY} \label{killpot}
\feqn



The scalar potential 
can be expressed for a generic number of hypermultiplets and vector
multiplets as
\begin{equation}
{\cal V}= g^2 [- P_r P^r +2 P_{xr} P^r_y g^{xy} + 2 N_{Ai}N^{Ai}] \label{scalarpot}
\end{equation}
with \eqn & & N^{Ai}= \frac {\sqrt 6}4 h^I K_I^X f^{Ai}_X =
\frac 2{\sqrt 6} f^{Ai}_X R^{rYX} D_Y P^r \ , \\
& & P^r_x \equiv -\sqrt{\frac 32} \partial_x P^r = h_x^I P_I^r \feqn

Defining the superpotential $W$ by $P^r= \sqrt{\frac 32} W Q^r$ with $Q^rQ_r=1$ the potential becomes \eqn {\cal
V} = -6 g^2 W^2  + \frac92 g^2 [ g^{\Lambda \Sigma} \partial_\Lambda W \partial_\Sigma W + W^2 g^{xy}\partial_x
Q^r \partial_y Q_r ] \feqn where $\Lambda$ is the curl index of the entire $n_V+4n_H$--dimensional scalar
manifold. From the above relation it follows that the requirement on ${\cal V}$ to be of the form ${\cal V} = -6
g^2 W^2  + \frac92 g^2 [ g^{\Lambda \Sigma} \partial_\Lambda W \partial_\Sigma W]$, which ensures the
gravitational stability, is
$$\partial_x Q^r=0$$ as found in the sect. \ref{gauequa_v}.

The universal hypermultiplet ($n_H=1$) corresponds to the quaternionic K\"ahler space
$\frac{SU(2,1)}{SU(2)\times U(1)}$. A significant parametrization, from a M-theory point of view, is
\cite{ceresole:2001}
$$q^X= \{V,\sigma, \theta, \tau\}$$ with the metric
\begin{equation}
\rmd  s^2 = \frac{\rmd V^2}{2V^2} + \frac{1}{2V^2}\left( \rmd \sigma + 2 \theta \, \rmd  \tau - 2 \tau \, \rmd
\theta\right)^2 + \frac{2}{V} \, \left(\rmd \tau^2 + \rmd  \theta^2\right) \,. \label{quatmetric}
\end{equation}
Using the general properties of quaternionic geometry it is possible from (\ref{quatmetric}) to derive
explicitly all the quantities presented above, in particular the Killing vectors and the prepotentials of the
eight isometries of manifold. For the axionic shift we have:
\begin{equation}
\vec{k}_{(1)} = \left(
\begin{array}{c}
0 \\ 1 \\ 0 \\ 0
\end{array}
\right)\, ~~~~~~~~~~~~~~~~~~~~~~~~~~~~~~~~~~~~~ \vec{P}_{(1)} = \left(
\begin{array}{c}  0 \\ 0  \\ -\frac{1}{4V} \end{array}
\right)\,
\end{equation}
For $k_{(2)}$ and $k_{(3)}$ we have

\begin{equation}
\vec{k}_{(2)} = \left(
\begin{array}{c}
0 \\ 2\theta \\ 0 \\ 1
\end{array}
\right)\, ~~~~~~~~~~~~~~~~~~~~~~~~~~~~~~~~~~~~~ \vec{P}_{(2)} = \left(
\begin{array}{c} -\frac{1}{\sqrt{V}} \\ 0  \\ -\frac{\theta}{V} \end{array}
\right)\, \end{equation} \begin{equation} \vec{k}_{(3)} = \left(
\begin{array}{c}
0 \\ -2\tau \\ 1 \\ 0
\end{array}
\right)\, ~~~~~~~~~~~~~~~~~~~~~~~~~~~~~~~~~~~~~ \vec{P}_{(3)} = \left(
\begin{array}{c}  0 \\ \frac{1}{\sqrt{V}}  \\ \frac{\tau}{V}\end{array}
\right)\,
\end{equation}

\section{Equations of motion}
\label{app:motion} The equations of motion of the lagrangian (\ref{lagr_v}) in the presence of hypermultiplets
and vector multiplets are
\eqn
-\RR_{\mu\nu} +a_{IJ} F^I_{\mu a} F_\nu^{Ja} +g_{XY} D_\mu q^X D_\nu q^Y +g_{x y}
\partial_\mu \phi^x \partial_\nu \phi^y -\frac 16 |F|^2 g_{\mu \nu} +\frac 23 \VV g_{\mu \nu} =0 \cr
&&\label{einstein_v} \feqn from which it follows in particular \eqn -\frac 32 R +\frac 14 |F|^2 +\frac 32 g_{XY}
D_\mu q^X D^\mu q^Y +\frac 32 g_{xy} \partial_\mu\phi^x \partial^\mu\phi^y - \frac 94 |a^I K_I|^2 + 5 \VV =0
\label{particular_v} \feqn
The variation with respect to the gauge fields gives \eqn \DD_a (a_{IK} F^{Kae})
+\frac 1{2 \sqrt 6} C_{IJK} \epsilon^{abcde} F^J_{ab} F^K_{cd} -g K^X_I D^e q^Y g_{XY}=0 \label{maxwell_v} \feqn
Finally the equations for the scalars are \eqn & & \hat{D}_\mu D^\mu q^W +gA^{\mu I} D_\mu K_I^W = g^{WX}
\partial_X
\VV \label{qq_v} \\
& & \hat{D}_\mu D^\mu \phi^x +gA^{\mu I} D_\mu K_I^x = g^{x y}
\partial_y \VV +\frac 14 g^{x y} \partial_y a_{IJ} F^I_{\mu \nu}
F^{J\mu \nu} \label{fifi_v} \feqn Here $\DD$ is the covariant derivative with respect to the spin connection and
$\hat{D}$ is a totally covariant derivative, ie with respect to all the indices. So for example \eqn \hat{D}_\mu
D^\mu \varphi^\Lambda =\DD_\mu D^\mu \varphi^\Lambda +\Gamma^\Lambda_{\Sigma\Theta} D_\mu \varphi^\Sigma D^\mu
\varphi^\Theta \nonumber \feqn and in general \eqn D_\mu f^* (q ,\phi )= D_\mu q^X \partial_X f^* + D_\mu \phi^x
\partial_x f^*
\feqn Now we specialize the above relations to the problem studied in this work. Due to symmetry of the class of
solutions considered only the Einstein equations for the components $(tt)$, $(rr)$ and $(\theta \theta)$ are
independent: \eqn & & -e^{v-w} \partial_r (v^\prime e^{v-w}) -3\frac {v^\prime}r  e^{2(v-w)} +e^{2(v-w)}
(v^\prime \Lambda + \Lambda^\prime)^2 + 4 g^2 e^{2v} W^2~~~~~~~~~~~~~~~~~ \cr & & ~~~~~~~- 3 g^2 e^{2v} (1- 3
\Lambda^2) [\frac 1{1-\Lambda^2} g^{xy} \partial_x W \partial_y W + g^{XY} \partial_X W \partial_Y W] = 0
\label{e-tt_v} \feqn

\eqn & & e^{w-v} \partial_r (v^\prime e^{v-w}) -\frac 3r w^\prime - (v^\prime \Lambda + \Lambda^\prime)^2 - 4
g^2 e^{2w} W^2 \cr & & ~~~~~~~~~~~~~~~~~~~~~~~+ 3 g^2 e^{2w} (1 + 3(1-\Lambda^2))[ \frac 1{1-\Lambda^2} g^{xy}
\partial_x W \partial_y W + g^{XY}
\partial_X W \partial_Y W] = 0
\cr
& &\label{e-rr_v} \\
& & re^{-2w} (v^\prime -w^\prime ) -2(1-e^{-2w}) + \frac 12 r^2 e^{-2w} (v^\prime \Lambda + \Lambda^\prime)^2 -
4 g^2 r^2 W^2 \cr & & ~~~~~~~~~~~~~~~~~~~~~~~+ 3 g^2 r^2 [\frac 1{1-\Lambda^2} g^{xy}
\partial_x W \partial_y W + g^{XY} \partial_X W \partial_Y W] = 0
\label{e-thetatheta_v} \\
& & \sqrt {\frac 32 }ge^v a^I K_{IX} q^{\prime X} =0 \label{e-rt_v} \feqn where we use the BPS relations for
$\phi^{\prime x}$, $q^{\prime X}$ and $|K|^2=6 g^{XY} \partial_X W \partial_Y W$. Following the manipulations of
\cite{noialtri} we consider the sum of (\ref{e-tt_v}) and (\ref{e-rr_v}) multiplied by $e^{2(v-w)}$ \eqn
\frac{(v^\prime + w^\prime)}r e^{-2w} & = & 3 g^2 [\frac 1{1-\Lambda^2} g^{xy} \partial_x W \partial_y W + g^{XY} \partial_X W \partial_Y W] \\
& = & \pm g \frac{e^{-w}}{\sqrt{1-\Lambda^2}} W^{\prime} \label{tt+rr_v} \feqn The above expression is the
direct generalization of the one in \cite{noialtri} and is identically satisfied by (\ref{1fr_v}) and
(\ref{b-c_v}). Now by the substitution of (\ref{tt+rr_v}) in (\ref{e-tt_v}), (\ref{e-rr_v}) and
(\ref{e-thetatheta_v}) it is easy to check that also these expressions are identically satisfied by the set of
BPS equations. Finally (\ref{e-rt_v}) is solved by (\ref{333_v}) and the equation $q^{\prime Z} K_Z =0$, which
follows for example from (\ref{erreeffe_v}) and (\ref{3restrizione_v}).

Next consider the equations for the gauge fields: \eqn
& & K^I_X q^{\prime X}=0 \label{em-r_v} \\
& & \partial_r (a_{IJ} e^{-w} r^3 (v' a^J +a^{\prime J})) -e^w r^3 g^2 g_{XY} K_I^X K_J^Y a^J =0 \label{em-t_v}
\feqn It is convenient to project these equations on the base $(h_I , h_J^x )$. The contraction of
(\ref{em-r_v}) with $h_I$ gives $K_X q^{\prime X} =0$ which we have already shown to be a consequence of BPS
equations.

The contraction with $h_{Ix}$ gives $q^{\prime X} K_X^I h_{Ix} =0$ which by means of (\ref{erreeffe_v}) is
equivalent to $\partial^X W K_X^I h_{Ix} =0$. But from (\ref{6restrizione_v}) and (\ref{chiq_v}) we have \eqn
\partial_x K^Z =\frac {\partial_x W}W K^Z  \label{simply_v}
\feqn so that \eqn
\partial^X W K_X^I h_{Ix} =\sqrt {\frac 32} \partial^X W \partial_x K_X
=\sqrt {\frac 32} \frac {\partial_x W}W \partial^X W K_X =0 \feqn After an integration by parts and using
(\ref{key_v}) the contraction of (\ref{em-t_v}) with $h^I$ gives

\eqn
\partial_r [r^3 e^{-w} (v' \Lambda +\Lambda')] +\sqrt {\frac 23} r^3 e^{-w}
\phi^{\prime x} h_{Ix} (v' a^I +a^{\prime I}) -e^w r^3 g^2 g_{XY} K^X K^Y_J a^J =0\cr && \feqn and using
(\ref{bene_v}) becomes
\eqn
\partial_r [r^3 e^{-w} (v' \Lambda +\Lambda')] \mp \frac {2gr^3 \Lambda
W'}{\sqrt {1-\Lambda^2}} \pm \frac {2gr^3 \Lambda q^{\prime X}
\partial_X W}{\sqrt {1-\Lambda^2}} -e^w r^3 g^2 g_{XY} K^X K^Y_J a^J =0 \cr
&& \feqn The above equation, after the use of (\ref{333_v}), can be easily related to the computations in
\cite{noialtri} with $\Lambda$ in place of $a$.

The contraction with $h_y^I$ gives \eqn & & r^3 e^{-w} (\Lambda' +v' \Lambda) \sqrt {\frac 23} \phi^{\prime x}
h_y^I h_{Ix} \mp h_y^I \partial_r h_I^x \frac {\partial_x W \sqrt 6 g\Lambda r^3}{\sqrt
{1-\Lambda^2}}~~~~~~~~~~~~~~~~~~~~~~~~~~~~~~ \cr & & ~~~~~~~~~~~~~~~~~~~~~~~\mp \partial_r \left( \frac
{\partial_y W \sqrt 6 g\Lambda r^3}{\sqrt {1-\Lambda^2}} \right) =-g^2 r^3 e^w \sqrt {\frac 32} \partial_y K^Z
K_Z \feqn From (\ref{1fr_v}) and (\ref{strana_v}) we find \eqn -rW \phi^{\prime x} =3\partial^x W \feqn and from
(\ref{corol_v}), (\ref{1fr_v}) \eqn g^2 e^{2w} rW =-2(W' -\phi^{\prime x} \partial_x W) \feqn Using these last
equations together with (\ref{simply_v}) and (\ref{b-c_v}) we have \eqn h_y^I \partial_r h_I^x \frac {\partial_x
W \Lambda r^2}W +\Lambda \partial_r \left( \frac {\partial_y W r^2}W \right) =-r^2 \Lambda \frac {\partial_y W}W
\phi^{\prime x} \frac {\partial_x W}W  \label{chefatica_v} \feqn By means of (\ref{chiq_v}) and some integration
by parts the following identity can be derived
\eqn h_y^I \partial_r h^x_I \partial_x W &=& \sqrt {\frac 23}
h_y^I
\partial_r h_I^x
\partial_x h^J P_J^s Q_s \cr &=& -h_y^I \partial_r (h_I^x h_x^J )\frac 23 P^s_J Q_s - \sqrt {\frac 23}
\partial_r
\partial_y h^J P^s_J Q_s \cr &=& h_y^I \partial_r (h_I h^J) \frac 23 P_J^s Q_s -\sqrt {\frac 23} \partial_r
[\partial_y (P^s Q_s)] +\sqrt {\frac 23} q^{\prime X} \partial_X [\partial_y (P^s Q_s)] \cr &=& \frac 23 g_{yx}
\phi^{\prime x} W -\partial_r (\partial_y W) +q^{\prime X}
\partial_X (\partial_y W) \cr
&=& -2r\Lambda \frac {\partial_x W}W-\partial_r (\partial_y W) +q^{\prime X}
\partial_X (\partial_y W) \label{fondam_v}
\feqn where in the last step we have used (\ref{strana_v}) and (\ref{1fr_v}). This together with
(\ref{erreeffe_v}) and (\ref{7restrizione_v}) shows that (\ref{chefatica_v}) is identically satisfied.

The equations of motion for the hyperini are \eqn & &e^{-(v+w)} r^{-3} \partial_r (r^3 e^{v-w} g_{ZY} q^{\prime
Y}) -\frac 12 q^{\prime X} \partial_X g_{ZY} q^{\prime Y} e^{-2w} +\frac 34 g^2
\partial_Z g_{XY} a^I K_I^X a^J K_J^Y \cr
& & +\frac 32 g^2 g_{XY} a^I a^J \partial_Z K_I^X K_J^Y = g^2 \partial_Z \left( -6W^2 +\frac 34 K^2 +\frac 92
g^{xy} \partial_x W
\partial_y W \right)
\feqn Using (\ref{333_v}), (\ref{erreeffe_v}), (\ref{7restrizione_v}) and (\ref{strana_v}) it becomes \eqn \pm
e^{-w} 9 \frac gr \sqrt {1-\Lambda^2} \partial_Z W &\pm& 3(v'-w') ge^{-w} \sqrt {1-\Lambda^2} \partial_Z W \cr &
\pm& e^{-2w} 3g \partial_Z W \partial_r (e^w \sqrt {1-\Lambda^2}) =-12g^2 W\partial_Z W \cr & & \feqn which
follows from (\ref{1fr_v}) and the considerations in \cite{noialtri}.

The equations of motion for the gaugini are \eqn & &\frac 34 \partial_x a_{IJ} e^{-2w} (a^{\prime I} +v' a^I)
(a^{\prime J} +v' a^J) -\frac 12 \partial_x g_{yz} \phi^{\prime y} \phi^{\prime z} e^{-2w} +r^{-3} e^{-(v+w)}
\partial_r (r^3 e^{v-w} \phi^{\prime y} g_{xy}) \cr & & -g^2 \partial_x \left( -6W^2 +\frac 34 K^2 +\frac 92
g^{zy} \partial_z W
\partial_y W \right)=0
\feqn From (\ref{strana_v}), (\ref{erreeffe_v}) and (\ref{7restrizione_v}) we find \eqn & & r^{-3} e^{-(v+w)}
\partial_r (r^3 e^{v-w} \phi^{\prime y} g_{xy}) = \pm 9\frac gr \frac {e^{-w} \partial_x W}{\sqrt {1-\Lambda^2}}
\pm 3gv' \frac {e^{-w} \partial_x W}{\sqrt {1-\Lambda^2}} \pm 3g e^{-w} \partial_x W \partial_r \frac 1{\sqrt
{1-\Lambda^2}} \cr & &~~~~~~~~~~~~~~~~~~~~~~~~~~~~~~~~~~~~ +3g \frac {e^{-w}}{\sqrt {1-\Lambda^2}} \left[ \frac
{3ge^{w} \partial_y\partial_x W \partial^y W}{\sqrt {1-\Lambda^2}} +3ge^w \partial_x \partial_X W \partial^X W
\sqrt {1-\Lambda^2} \right] \cr &&
\\
&& -g^2 \partial_x \left( -6W^2 +\frac 34 K^2 +\frac 92 g^{zy} \partial_z W
\partial_y W \right) = 12g^2 W\partial_x W -9g^2 \partial_x \partial_Y W
\partial^Y W \cr
&&~~~~~~~~~~~~~~~~~~~~~~~~~~~~~~~~~~~~~~~~~~~~~~~~~~~~~~ -  \frac 92 g^2 \left(\partial_x g^{yz} \partial_y W \partial_z W + 2 \partial_x \partial_y W \partial^y W\right)\\
& & -\frac 12 \partial_x g_{yz} \phi^{\prime y} \phi^{\prime z} e^{-2w} = -\frac 92 g^2 \frac {\partial_y W
\partial^y W}{\sqrt {1-\Lambda^2}} \label{elaborate_v} \feqn In a similar way as  (\ref{fondam_v}) one finds
\eqn
\partial_x h_I^y \partial_y W =\frac 23 W h_{Ix} +h_I \sqrt {\frac 23}
\partial_x W -h_I^y \partial_x \partial_y W
\feqn From this and (\ref{key_v}), (\ref{bene_v}) and (\ref{a_v}) we find \eqn & & \frac 34 \partial_x a_{IJ}
e^{-2w} (a^{\prime I} +v' a^I) (a^{\prime J}+v' a^J) =\mp \frac {6ge^{-w} \partial_x W}{\sqrt {1-\Lambda^2}}
\left[ v'-\frac {1-\Lambda^2}r \right] +\frac {6\Lambda^2 g^2 W\partial_x W}{1-\Lambda^2} \cr & &
~~~~~~~~~~~~~~~~~~~~~~~~~~~~~~~~~~~~~~~~~~~~~~~-\frac 92 g^2 \frac{\Lambda^2}{1-\Lambda^2} \left(\partial_x
\partial_y W \partial^y W -2 \partial_x g_{yz} \partial^y W \partial^z W\right)\cr && \feqn Summing up all the
terms \eqn 0 &=& \left( \pm 9\frac gr e^{-w} \mp 3ge^{-w} v' \pm 3ge^{-w} \frac {\Lambda \Lambda'}{1-\Lambda^2}
\right) \frac {\partial_x W}{\sqrt {1-\Lambda^2}} +12 g^2 W\partial_x W \cr &+& 6\Lambda^2 g^2 \frac
{W\partial_x W}{1-\Lambda^2} \pm 6ge^{-w} \partial_x W \sqrt {1-\Lambda^2} \feqn Using (\ref{1fr_v}) to
eliminate $e^{-w}$ from the first and the last term and next (\ref{a_v}) to eliminate $\Lambda' +\Lambda v'$, we
finally see that the gaugini equations also are satisfied.
\end{appendix}

\end{document}